# Multilingual Medical Documents Classification Based on MesH Domain Ontology

**Elberrichi Zakaria    Taibi Malika    Belaggoun Amel**

**Computer Science Department, University Djillali Liabes, EEDIS Laboratory**
**Sidi Belabbes, 22000, Algeria**

**Abstract**
This article deals with the semantic Web and ontologies. It addresses the issue of the classification of multilingual Web documents, based on domain ontology. The objective is being able, using a model, to classify documents in different languages. We will try to solve this problematic using two different approaches. The two approaches will have two elementary stages: the creation of the model using machine learning algorithms on a labeled corpus, then the classification of documents after detecting their languages and mapping their terms into the concepts of the language of reference (English). But each one will deal with the multilingualism with a different approach. One supposes the ontology is monolingual, whereas the other considers it multilingual.  To show the feasibility and the importance of our work, we implemented it on a domain that attracts nowadays a lot of attention from the data mining community: the biomedical domain. The selected documents are from the biomedical benchmark corpus Ohsumed, and the associated ontology is the thesaurus MeSH (Medical Subject Headings). The main idea in our work is a new document representation, the masterpiece of all good classification, based on concept. The experimental results show that the recommended ideas are promising.
***Keywords:*** *multilingual classification, medical document, concept, domain's ontology, Ohsumed, MeSH*

## 1. Introduction

The exponential growth of information on the Internet generates the difficulty in finding, and organizing this information disseminated in the four corners of the planet, and implies a de facto multilingualism. Thus, techniques and tools are needed to automatically access the relevant information, beyond language barriers. These various tools, which constitute a major element of the Semantic Web, however require a particular formalization of the contents, by the addition of a semantic description, generally carried out by metadata.

Ontologies, one of the most used models of knowledge representation, answer, as we will try to prove it, these problems. They organize knowledge according to the application domain considered and consist of concepts bound by relations.

The remainder of this document is structured as follows: some related works are presented in chapter 2. The architectures of our approaches are presented in section 3 with their various stages. The evaluations of these approaches are in section 4. And at the end, a conclusion and future works are in section 5.

## 2. Related works

Document representation for classification is typically based on the traditional approach `Bag-of-words'. However, during the last few years, researchers of the field of text mining, have tried to improve this conventional representation. One promising way being the use of concepts. One of these improved approaches implies the use of ontologies.

The representation of text as a bag of words was prejudiced by the ignorance of any relation between these terms, thus the importance of amine's work [3] who tried to integrate an ontology (Wordnet) to improve the process of textual documents clustering.

Jalam and Clesh propose in [1], three approaches for the categorization of multilingual texts, which are founded on the translation of the documents towards a reference language.

Guyot et al. in [5] propose an approach which uses a multilingual ontology for information search, without using translation. They only tried to prove the feasibility of the approach. Nevertheless, limits exist because of incomplete ontology used.

Sanchez and al. in [2] propose the opposite of the extraction of information using an ontology. They use Web documents to create ontology based on statistical and linguistic methods (Wordnet) in a given field. The basic idea being: a good design of ontology implies strong semantics, which implies a pertinent and an appropriate Web documents classification.

Litvak and al. in [6] propose a method of classification of the multilingual documents Web by using a multilingual ontology for the conceptual representation of the documents.

Song and al. in [7] concretize an approach consisting in classifying Web documents automatically by using domain ontology, without using neither learning algorithms, nor a learning base.

## 3. The Proposed Approaches

A problem to be solved in the documents (texts) classification is: How to represent the texts in order to facilitate their processing, and especially how to preserve only information useful for the classification. The representation most largely used in this field is the "bag of words" representation.

Many works were proposed to lessen the disadvantages of this representation [4]. In our approaches, we propose a method that combines the words with their associated concepts. We will present thereafter two ways of classifying multilingual Web documents, based on this new conceptual representation method. Although this representation is also based on the vectorial formalism to represent the documents, it remains basically different from the saltonian representation, as we will show it thereafter. Each approach will be particularized by its strategy of managing the multilingualism. Our approaches remain confronted with many problems even if they are better dealt with, compared to the traditional approach (bag of words):

- The processing time, because the number of the terms intervenes in the expression of the complexity of the algorithm; more this number is high more the computing time is important.
- The weak frequency of certain terms: we cannot build reliable classifiers starting from few occurrences in the training set. Also let us note that paradoxically, it was observed that the most frequent terms do not bring discriminating information since they are present everywhere.
- More the number of documents per category is high, more the memory capacity and the time computing is important, but on the other hand, more the developed model will be reliable and interesting.
- Certain of the existing methods (KNN for example) are based on the comparison between the document to classify and all the already classified documents. Therefore, more the number of these documents is important, more the response time will be high.
- The documents are generally of random sizes, and the size influences on the performances of the classification. Indeed, there exist methods which support documents of bigger sizes as there exist those which support documents of smaller sizes.
- The multilingual MeSH version under XML format is not easily exploitable on a machine Core2Duo 4G RAM , even say impossible using the INSERM 2011 edition .
- The French and English MeSH versions are not aligned one hundred percent: The design of a multilingual MeSH version was not easy. We were forced to add the French concepts manually to our monolingual ontology (English) to make it multilingual, because the French version MeSH (INSERM) was not exploitable. We supplemented the missing concepts in the French version by a translation of the English concepts using an API GOOGLE supervised by domain expert.
- MeSH contains 16 subcategories; we worked with the subcategory C for diseases, which corresponds well to the selected diseases corpus Ohsumed.

### 3.1 A conceptual document representation

As we already said, the representation mode of the document plays a key role in automatic classification, and has direct effect on its performance. We choose to perform a conceptual representation on our medical document.

So, a conceptualization step, a particularity of our approach, is applied, using a mapping strategy. The process of mapping terms into concepts is illustrated with an example shown in Figure 1.

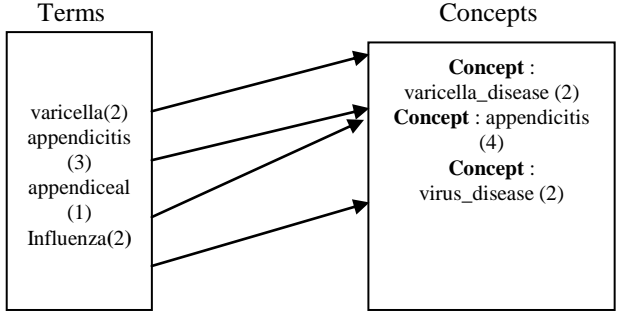

Fig. 1- Example of mapping terms into concepts

At this point, three strategies of mapping can be distinguished:

**Add Concept:** This strategy consists in extending each vector of terms $t_d$ with new entries of the concepts that appear in the text. Thus the vector $t_d$ will be replaced by the concatenation of the vector $t_d$ with the vector $c_d$ where $c_d = (cf(d,c_1)\ldots cf(d,c_k))$ Represents the vector of the concepts with K =|C| and cf(d, c) represents the frequency of appearance of the concept c ∈C in category d.

The terms of the vector $t_d$ which appear in the ontology will be duplicated in the new representation (once in the old terms vector and once the concepts vector).

**Replace the terms by concepts:** This strategy is similar to the first; the only difference lies in the fact that it avoids the duplication of the terms in the new representation i.e. the terms which appear in ontology will be present only in the concepts vector. The vector of the terms will thus contain only the terms which do not appear in ontology.

**Concept only:** This strategy differs from the second by the fact that it excludes all the terms from the new representation including the terms which do not appear in the ontology. $c_d$ is used to represent the document.

**Strategies of disambiguation:** It is clear that assigning a term to its concepts in an ontology is neither simple, nor direct because of the natural ambiguity of the language. For this reason to add or replace terms by concepts can imply a risk in certain cases, of a loss of information. The choice of the more appropriate concept for a term will influence logically on the performances of the classification. We distinguish two strategies for disambiguation:

**All concepts**: This strategy is the simplest because it quite simply consists in taking all the concepts suggested by the ontology, but it will result in the increase the representation space. The frequencies of concepts will be calculated as follows:

$$cf(d,c) = tf\{d, t \in T | c \in (ref_c(t))\} \quad (1)$$

**First concept**: If the ontology used gives for each term an ordered list of concept according to a certain criterion. This strategy of disambiguation consists in taking only the first concept of the list as the most suitable concept. The frequencies of concepts will thus be calculated as follows:

$$cf(d,c) = tf\{d, t \in T | first(ref_c(t)) = c\} \quad (2)$$

**Hyperonymy:** After having decided on the strategy of mapping as well as the disambiguation strategy, the next stage consists in enriching the representation by including the hyperonyms. So, the frequencies in the vector of the concepts will be updated as follows:

$$cf'(d,c) = \sum_{b \in H(c)} cf(d,b) \quad (3)$$

Where H (c) represents the set of the Hyperonyms concept C.

In our approach, we chose to experiment with two possibilities: first, the mapping strategy that replace term by concept and the taking the first concept disambiguation and second enriching the representation vector with hyperonyms.

2.2 The learning phase

The data used for classification are Web documents that contain html tags and images or other noise sources to be deleted, leaving only place to text. At this level, if the words are preserved as is, the machine will not recognize them automatically as the same words in their various versions. That's one of the reason, we have a preprocessing step of the texts.

**Preprocessing :** The use of words and stemma for the representation of texts requires preprocessing to make the classification as efficient as possible, and for a better relevance of the information. Indeed, in textual documents many words provide little (if any) information on the relevance of the document. Algorithms called stop words removers specialize in eliminating them. That will allow a reduction in the representation space and in the processing time.

In our case a language detection step is necessary, because our base is multilingual; this step is automatic according to the stop words contained in each document. After the detection of the language, the stops words will be removed and a vectorization is applied for each document; a French vector or an English vector.

Our key contributions in this paper are promising and original management of multilingualism for medical documents classification. Multilingualism in our case is limited to two languages, English and French. It can be generalized to other language without any problems for the approach, provided you adapt it and have the corresponding multilingual ontology MeSH. English as a universal renowned language will serve as the pivot language, the one that will be used to design the model.

**Approach by translation:** This step is applied only on the French vectors, since it considers English the main language, and French (or other languages) as secondary in this situation and in our approaches to multilingualism. We distinguish two approaches, one of them, we called the translation approach.

The translation of texts is a difficult task and is never perfect. To minimize the risks, in our approach, we translate the terms vectors and not the text. For that we will use Google Api translate Java which provides an automatic translation for many languages (English, French, Russian, Japanese, Arabic, ... etc).

This approach consist in translating the French term vector into a English term vector, then continue the process (conceptualization) as if it was an English term vector using the English mesh ontology to have English concept vector. In a second phase, we will enrich the representativeness by adding the hyperonyms. Strategies of mapping and disambiguation were explained in the conceptual document representation section.

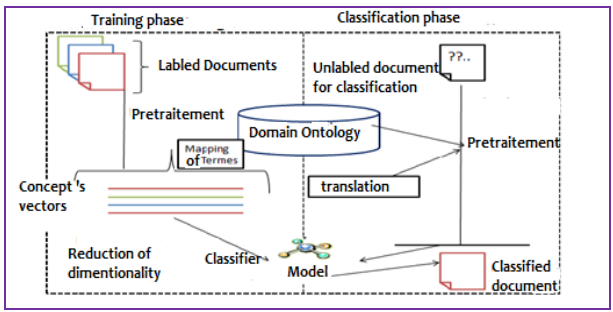

Fig.2 Approach by translation

**Approach by Multilingual Ontology :** We do things differently in this approach. To manage the multilingualism, in this approach, we begin by mapping the French terms vector (or another language) in the French concepts using the French part of the ontology (since it is multilingual), and then we look for the English concepts that correspond to the French ones, to have our English concept vector. In a second phase, we will enrich the representativeness by adding the hyperonyms. Strategies of mapping and disambiguation were explained in the conceptual document representation section.

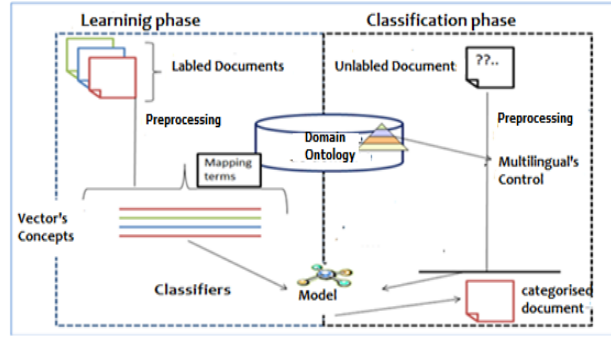

Fig.3 Approach by multilingual MeSH

### 3.3 Weighting

Weighting makes it possible to represent the importance of the term in a category. The number of occurrences of the term in the category is the simplest way to calculate this value but it is not very satisfactory, because it does not take into account the other categories.

The measure most widely used is known as the TF-IDF, it was introduced within the framework of the vectorial model, Its initials mean:
"Term Frequency" * "Inverse Document Frequency"
"TF: Term Frequency" it is quite simply the number of occurrences of the term in the category concerned. The "IDF: Inverse Document Frequency" is the reverse of the total number of category divided by the number of categories containing the term, nbr-category is the number of categories is the formula:

$$TFIDF(c_i, w_j) = TF(c_i, w_j) * \log\left(\frac{nbre - categorie}{DF(w_j)}\right) \quad (4)$$

### 3.4 The classification phase

As it is always done in machine learning, to classify a document, we expose the vector of descriptors of the document to be classified on the model created in the learning stage, in order to determine the adequate class of the new document. Of course, the document will be processed in the same way as the other documents of the learning phase as explained in the conceptual document representation section.

# 4. The Evaluation of the Approaches

We evaluate our approaches using the most popular algorithms of classification like Naïve Bayes, C4.5, KNN, and AdaboostM1

4.1 Collection test

We use the OHSUMED collection, proposed within the framework of the TREC9-Filtering task in 2000, which consists of the titles and/or summaries of 270 medical journals published between 1987-1991 (Hersh and Al 1994). A document contains six fields: title (. T), absstract (. W), MeSH indexed concepts (. M), author (. A), source (. S), and publication (. P).

Table 1: Distribution of the documents in the Ohsumed corpus categories

| Categories | # Docs |
|---|---|
| **Bacterial Infections and Mycoses** | 2540 |
| **Virus Diseases** | 1171 |
| Parasitic Diseases | 427 |
| Neoplasms | 6327 |
| Musculoskeletal Diseases | 1678 |
| **Digestive System Diseases** | 2989 |
| **Stomatognathic Diseases** | 526 |
| **Respiratory Tract Diseases** | 2589 |
| Otorhinolaryngologic Diseases | 715 |
| **Nervous System Diseases** | 3851 |
| **Eye Diseases** | 998 |
| Urologic and Male Genital Diseases | 2518 |
| Female Genital Diseases and Pregnancy Complications | 1623 |
| Cardiovascular Diseases | 6102 |
| Hemic and Lymphatic Diseases | 1277 |
| Neonatal Diseases and Abnormalities | 1086 |
| Skin and Connective Tissue Diseases | 1617 |
| Nutritional and Metabolic Diseases | 1919 |
| Endocrine Diseases | 865 |
| Immunologic Diseases | 3116 |
| Disorders of Environmental Origin | 2933 |
| **Animal Diseases** | 506 |
| Pathological Conditions, Signs and Symptoms | 9611 |

In order to obtain for our approach, a multilingual Ohsumed version, we added medical French documents in each category of the Ohsumed corpus.

4.2 MeSH ontology

We used the referenced thesaurus of the biomedical domain developed by the NLM (National Library of Medicine) in the United States. The thesaurus MeSH (Medical Subject Headings) is a tool used for indexing and searching for medical information. The first version appeared in 1954 named Subject Heading Authority List. Its publication under the name of Medical Subject Headings goes back to 1963 and it contained, in this edition, 5.700 descriptors. Since, MeSH did not cease growing. It counts 25588 descriptors in its 2010 version. MeSH descriptors are organized in 16 categories: the category A for anatomical terms, the category B for organisms, the category C for diseases, etc. Each category is subdivided in subcategories. Inside each category, the descriptors are structured hierarchically, from the most general to the most specific, with a maximum depth level of 11.

The MeSH thesaurus is used as a controlled vocabulary for the indexing of the resources of the bibliographical database MEDLINE20. It is also used also by the portals for indexing and cataloguing medical resources, such as Health On the Net and CisMef. The INSERM maintains a French version of MeSH.

4.3 Criteria of evaluation

To evaluate our algorithms, and our approach, we used the most used measure by the supervised classification community, "the measure F", which is the harmonic average of the precision and the recall.

$$F = \frac{2 * recall * precision}{recall + precesion} \quad (5)$$

In the formula above, the precision and the recall are two standard measurements largely used in the process of classification to evaluate the effectiveness of the algorithms on a given category.

$$precision = \frac{true\ positive}{true\ positive + false\ positive} \quad (6)$$

$$recall = \frac{true\ positive}{true\ positive + false\ negative} \quad (7)$$

4.4. Experimental Results

To be able to show the utility of the use of The Mesh domain ontology in multilingual classification in a conceptual representation, we tested the approach suggested on our corpus, the multilingual Ohsumed.

The multilingual base was tested on both approaches according to several criteria: the number of the categories, as the algorithms of classification, for both representations conceptual and conceptual with hyperonyms.

The following tables summarize the results for the two approaches translation and multilingual Ontology Mesh for the multilingual Ohsumed corpus.

Table 2: Comparison of F-measures for 8 categories

| Representation | | Concepts | | | Concepts +Hyperonyms | | |
|---|---|---|---|---|---|---|---|
| Algorithms | | KNN | C4.5 | Ada BoostM1/NB | KNN | C4.5 | Ada BoostM1/NB |
| Mac Avg | MeSH Multiling | 0.80 | 0.706 | 0.77 | 0.813 | 0.808 | **0.83** |
| | Translate | 0.791 | 0.706 | 0.785 | 0.791 | 0.733 | **0.79** |

Table 3: Comparison of F-measures for the whole Corpus

| Representation | | Concepts | | | Concepts +Hyperonyms | | |
|---|---|---|---|---|---|---|---|
| Algorithms | | KNN | C4.5 | Ada Boost M1/NB | KNN | C4.5 | Ada Boost M1/NB |
| Mac avg | MeSH Multiling | **0.816** | 0.80 | 0.803 | 0.704 | 0.700 | 0.701 |
| | Translate | 0.791 | 0.734 | 0.791 | **0.801** | 0.734 | 0.797 |

Studying these results, we can affirm that multilingual classification can be dealt with without any problems using a conceptual ontology, and gives good results. Both approaches are good, and depending on various factors (the algorithms, the representation) perform better than the other. The most important we proved the viability and the importance of both approaches. The results are heterogeneous and very interesting for a multilingual environment, and reach 83%.

Given these results we can say that the classification based multilingual compared to the translation approach. We are not comfortable to experiment these two approaches in a single context, so for greater certainty, we varied the context. Also, we used two conceptual representations, with and without hypernyms, and several data mining algorithms, the K nearest neighbors, decision trees, and the metalearning algorithm Adaboost applied to Naives bayes.

And finally, we have done all that on two occasions. Once on a small corpus, composed of only eight categories (those in bold in Table 1). Then again, on the whole corpus benchmark Ohsumed.

Therefore, the results can have various readings, we encourage the informed reader to make their own from our tables 2 and 3.

The KNN algorithm and Naives Bayes algorithm used in the AdaboostM1 have given good results, as always, when used for classification of textual documents. This is what has justified our choice of these algorithms.

## 5. Conclusions

The main goal of our approach is to improve the process of multilingual medical document classification based on domain ontology; our method was tested using the medical domain ontology MesH. Accordingly a conceptual document representation was used.

The experiments carried out led to the following observations:
- Domain ontology can be efficiently used for document classification.
- Conceptual representation is a promising way to go, especially since the web semantic is the ultimate goal.
- Multilingual document classification is the future of document classification, but the good news is, it can be solved in many ways. We just proposed two ways for your judgment.

At the conclusion of this work, we think many tracks remain to be exploited. The first prospect relates to the addition of other languages in the MeSH ontology by translating the concepts in a supervised way (guided by an expert in the medical field). Another prospect consists in going up more than only one level in the MeSH ontology (Hyperonyms); we could try to generalize to a maximum and have by the way reduced calculations. Concerning the reduction of dimensionality; we propose to use techniques like the LSI.

**Zakaria Elberrichi** received his Master degree with thesis in computer science from the California State University, in addition to PGCert in higher education, and received his PhD in computer science from the university Djillali Liabes, Sidi-Belabbes, where he has been a faculty member ever since. He is also a member of EEDIS (Evolutionary Engineering and Distributed Information Systems) laboratory and the project head - manager of the data mining and intelligent web team.

**Taibi Malika** and **Belaggoun Amel** received their Master with thesis degree from the computer science department at the UDL university with honor in 2011, and are currently doctorate students and member of the research team intelligent web mining.